\documentclass{SCIS2019}
\usepackage{mathtools}        
\usepackage{amssymb}
\usepackage{graphicx}
\usepackage{graphics}
\usepackage{verbatim}
\usepackage[colorlinks]{hyperref}

\newcommand{\tr}{\mathrm{Tr}}

\begin{document}
\ArticleType{RESEARCH PAPER}

\Year{2019}
\Month{May}
\Vol{}
\No{}
\DOI{}
\ArtNo{}
\ReceiveDate{}
\ReviseDate{}
\AcceptDate{}
\OnlineDate{}

\title{On Robust Spectrum Sensing Using M-estimators of Covariance Matrix}{A Robust Spectrum Sensing Using M-estimators of Covariance Matrix}

\author[1]{Zhedong Liu}{}
\author[1]{Abla Kammoun}{}
\author[1]{Mohamed Slim Alouini}{}

\AuthorMark{Zhedong Liu}

\AuthorCitation{Zhedong Liu, Abla Kammoun, Mohamed Slim Alouini}


\address[1]{King Abdullah University of Science and Technology , Thuwal {\rm 23955}, Saudi Arabia}

\abstract{In this paper, we consider the spectrum sensing in cognitive radio networks when the impulsive noise appears. We propose a class of blind and robust detectors using M-estimators \textcolor{blue}{in eigenvalue based spectrum sensing method. The conventional eigenvalue based method uses statistics derived from the eigenvalues of sample covariance matrix(SCM) as testing statistics, which are inefficient and unstable in the impulsive noise environment. Instead of SCM, we can use M-estimators, which have good performance under both impulsive and non-impulsive noise.} Among those M-estimators, We recommend the Tyler's M-estimator instead, which requires no knowledge of noise distribution and have the same probability of false alarm under different \textcolor{blue}{complex elliptically symmetric} distributions. In addition, it performs better than the detector using sample covariance matrix when the noise is highly impulsive. It should be emphasized that this detector does not require knowledge of noise power which is required by the energy detection based methods. Simulations show that it performs better than conventional detector using sample covariance matrix in a highly impulsive noise environment.}

\maketitle
\section{Introduction}
The wide application of wireless network has stirred up a tremendous demand for bandwidth. Cognitive networks~\cite{cognitivesurvey,wang2014spatial} have been proposed as a promising solution to solve the problem of spectrum scarcity by making full use of the available spectrum. The users of Cognitive networks, or secondary users (SUs), have to be able to sense the free spectrum in which no signal of the licensed users, or primary users (PUs), exists. There are several techniques for the spectrum sensing, such as eigenvalue based spectrum sensing~\cite{cosensingRMTGLRT,cosensingRMTeigen,cosensingRMT}, energy detection\cite{urkowitz1967energy}, the matched filter\cite{poor2013introduction}, the cyclostationary feature detection\cite{enserink1994cyclostationary} and so on. Unfortunately, most of these techniques require knowledge of signal features of PUs or noise power. Among these techniques, eigenvalue based spectrum sensing requires no information about both signal and noise and only a few numbers of samples. Thus eigenvalue based spectrum sensing is the only method to fulfill all the stringent requirements and limitations of the problem of spectrum sensing in the context of cognitive radio networks\cite{cosensingRMT}. 

Most of the sensing techniques are designed for Gaussian noise. The Gaussian assumption is always justified by central limit theorem, but these techniques do not deal with the non-Gaussian (impulsive or heavy-tailed) noise environment. In the wireless system, impulsive (heavy-tailed) noise frequently occurs and originate from numerous sources, for instance, switching transients in power lines \cite{middleton1999non}, vehicle ignition \cite{batur2008measurements}, microwave ovens\cite{taher2008microwave} and devices with electromechanical switches\cite{blankenship1997measurements}. Under those circumstances, sensing techniques designed for Gaussian noise may be highly susceptible to severe degradation of performance.  Some existing detectors are designed to address the problem of spectrum sensing in impulsive noise environments \cite{lunden2010robust,kang2010class,moghimi2011adaptive,wimala2011}
. In \cite{lunden2010robust}, a robust detector is proposed based on the cyclic correlation detector. This work considered the symmetric $\alpha$-stable distribution. However, this detector requires cyclic frequencies of the PUs in advance thus it is not blind. A class of spectrum-sensing sensing schemes was proposed in \cite{kang2010class}. This detector uses the generalized likelihood ratio test at each antenna branch and combines them in a nonlinear way. However, this detector requires to know the exact noise distribution which is usually unavailable in many applications, e.g. cognitive radio networks. In \cite{moghimi2011adaptive}, a suboptimal $l_p$-norm detector was proposed. This method requires knowledge of the power of the fading channel gain and noise power. Also, the optimization of free parameters requires a large size of signal sample. Most of the existing detectors requires the knowledge of noise type either by prior knowledge or learn it from samples. Prior knowledge is usually unavailable, and, in high dimensional cases, learning the distribution non-parametrically requires a tremendous sample size which is also unavailable. 

In this paper, we propose a new spectrum sensing method to deal with the problem of non-Gaussian noise environment with limited information. The new method applies robust estimators of the covariance matrix  [17] to eigenvalue based spectrum sensing. \textcolor{blue}{The eigenvalue based spectrum sensing method detects the signal by exploiting the fact that the largest eigenvalue of the population covariance matrix of the received signal is greater than it is in the case of pure noise when the signal appears with certain signal to noise ratio. Then the task is simplified to estimate the population covariance matrix. Towards this goal, one natural approach consists in using sample covariance matrix(SCM), which has very bad performance in the impulsive noise environment. To improve the performance, we can use robust estimator instead of SCM. Specificly, we recommend to use Tyler's M-estimator.} When the detector uses Tyler's M-estimator, it becomes totally blind because it requires no information about signals and noise. It should be emphasized that this detector is distribution-free ,which means the noise type can be unknown and the detector will not need the noise distributions. As best of our knowledge, no detector has this distribution-free property. The robust estimator has 'good' performance in many noise environment, especially in complex elliptical symmetric distributed noise environment even though it is not optimal in general. The contributions and novelty of this paper are given as follows: 1) A novel spectrum sensing method is proposed for spectrum sensing applications; 2) the performance characteristics of the new method is compared with the conventional eigenvalue-based method.

The remainder of the paper is organized as follows. In section 2, we introduce the robust estimators of the covariance matrix or generally scatter matrix. The proposed robust eigenvalue based spectrum sensing is presented in Section 3. The performance of the proposed method is shown in Section 4. Section 5 concludes the paper. 
\section{Robust Estimators of The Covariance}
\subsection{ML-estimators}

Consider a zero-mean data set $\boldsymbol{x}_1,\dots,\boldsymbol{x}_n \in \mathbb{C}^p$, whose covariance matrix exists. The SCM $\boldsymbol{S}=\frac{1}{n}\sum_{i=1}^n{\boldsymbol{x}_i\boldsymbol{x}_i^H}$ is the ML-estimator of the covariance matrix $\boldsymbol{\Sigma}$ if $\boldsymbol{x}_i$'s are i.i.d. random vectors from the zero mean complex $p$-variate Gaussian distribution, denoted by $\mathbb{C}N_p(0,\boldsymbol{\Sigma})$. The complex $p$-variate Gaussian distribution is a special case of a class of complex elliptically symmetric (CES) distribution \cite{ollila2012complex}. The CES random vector has a stochastic representation. If $\boldsymbol{x}$ is a CES random variable, 
\begin{equation} \label{CES}
\boldsymbol{x}=\sqrt{\gamma}\boldsymbol{\Sigma}^{\frac{1}{2}}\boldsymbol{u},
\end{equation}
where $\gamma$ is a real univariate random variable known as texture parameter used to model the impulsive feature, $\boldsymbol{\Sigma}^{\frac{1}{2}}$ is Cholesky decomposition of a scatter matrix, and $\boldsymbol{u}$ is a $p$-dimensional random vector with uniform distribution over a hypersphere. Here we discuss the ML-estimation of the multivariate CES distribution with location parameter zero and scatter matrix $\boldsymbol{\Sigma}$ with probability density function (pdf) as follows,
$$f(\boldsymbol{x};\boldsymbol{\Sigma})=|\boldsymbol{\Sigma}|^{-\frac{1}{2}}g\{\boldsymbol{x}^H\boldsymbol{\Sigma}^{-1}\boldsymbol{x}\},$$
where $g$ is a positive valued function such that $f$ integrates to one. If the covariance matrix, $E[\boldsymbol{x}\boldsymbol{x}^H]$, of CES distribution exits, it is proportional to $\boldsymbol{\Sigma}$. One can show that the ML-estimates $\boldsymbol{\hat{\Sigma}}$ of the scatter matrix $\boldsymbol{\Sigma}$ is the solution to the following equation\cite{maronna2018robust}:
\begin{equation} \label{MLestimator}
\hat{\boldsymbol{\Sigma}}=\frac{1}{n}\sum_{i=1}^n u(\boldsymbol{x}_i^H\boldsymbol{\hat{\Sigma}}^{-1}\boldsymbol{x}_i)\boldsymbol{x}_i\boldsymbol{x}_i^H,
\end{equation}
\textcolor{blue}{where $\rho:d \to -2\ln{g(d)}$, $u:d \to \rho'(d)$ and $d \in \mathbb{R}$.}

Given an initial positive definite hermitian estimate  $\boldsymbol{\Sigma_0}$, define 
\begin{equation} \label{iter1}
\boldsymbol{\hat{\Sigma}}_{m+1}\leftarrow\frac{1}{n}\sum_{i=1}^n u(\boldsymbol{x}_i^H\boldsymbol{\hat{\Sigma}}_m^{-1}\boldsymbol{x}_i)\boldsymbol{x}_i\boldsymbol{x}_i^H.
\end{equation}
It has been shown that the sequence $\boldsymbol{\hat{\Sigma}}_{m+1}$ converge to the unique solution $\boldsymbol{\hat{\Sigma}}$ of \eqref{MLestimator}  under mild regularity conditions\cite{maronna2018robust}. In a practical implementation of the iteration \eqref{iter}, the iteration is usually terminated when $\|\boldsymbol{I}-\boldsymbol{\Sigma}^{-1}_{m-1}\boldsymbol{\Sigma}_{m}\|<\epsilon$, where $\|*\|$ is some matrix norm and $\epsilon$ is  some predetermined tolerance level, for example $\epsilon =0.001$. 
\subsection{M-estimators}
M-estimators of scatter matrix is a generalization of ML-estimators of scatter matrix for CES data. M-estimators are first introduced by Maronna\cite{maronna1976robust} ,and then Kent and Tyler proposed a more restricted class of redescending M-estimators \cite{kent1991redescending}. Both of them are studying the real case only, but it is very natural to extend M-estimators to complex data.

The M-estimators $\boldsymbol{\hat{\Sigma}}$ based on the data set $\boldsymbol{x}_1,\dots,\boldsymbol{x}_n \in \mathbb{C}^p$ is a solution to the following equation\cite{maronna2018robust}:
\begin{equation}\label{Mestimator}
\boldsymbol{\hat{\Sigma}}=\frac{1}{n}\sum_{i=1}^n u(\boldsymbol{x}_i^H\boldsymbol{\hat{\Sigma}}^{-1}\boldsymbol{x}_i)\boldsymbol{x}_i\boldsymbol{x}_i^H,
\end{equation}
where $u$ is a real valued function with certain requirement stated in \cite{maronna2018robust} ,and it is not necessarily derived from any CES distributions. The existence and uniqueness of $\boldsymbol{\hat{\Sigma}}$ is stated \cite{ollila2003robust} for complex data. The M-estimators can be interpreted as a weighted version of SCM whose weight is assigned by $u$ function. Here are some examples of the M-estimators.

{\bf SCM}. The SCM belongs to M-estimators since we can set \textcolor{blue}{$u(d)=1$}. This estimator is very sensitive to those extreme data points because the extreme data points share the same weights with the other data points. The SCM is also the ML-estimator for Complex Gaussian distribution.

{\bf Tyler's M-estimator}. Tyler's M-estimator is the solution to \eqref{Mestimator} with \textcolor{blue}{$$u(d)=\frac{p}{d}.$$} This estimator is also the ML-estimate of scatter for the complex angular central Gaussian distribution \cite{kent1997data}. If $n>p$ and $\boldsymbol{x}_i\neq\boldsymbol{0}$ for all $i$, \textcolor{blue}{given an initial positive definite hermitian estimate  $\boldsymbol{\Sigma_0}$, which can simply be the identity matrix,} this estimator can be computed by the iterations as follows.
\begin{equation} 
\boldsymbol{\hat{\Sigma}}_{m+1}\leftarrow\frac{p}{n}\sum_{i=1}^n \frac{\boldsymbol{x}_i\boldsymbol{x}_i^H}{\boldsymbol{x}_i^H\boldsymbol{\hat{\Sigma}}_m^{-1}\boldsymbol{x}_i}
\end{equation}
\begin{equation}\label{iter}
\boldsymbol{\hat{\Sigma}}_{m+1}\leftarrow \frac{\alpha\boldsymbol{\hat{\Sigma}}_{m+1}}{\tr \boldsymbol{\hat{\Sigma}}_{m+1}},
\end{equation}
where $\alpha$ a constant used to eliminate the scaling ambiguity and we can always set it to be $1$ or $p$.

{\bf ML-estimator of complex multivariate t-distribution}.
The ML-estimates $\boldsymbol{\hat{\Sigma}}$ of scatter $\boldsymbol{\Sigma}$ is the solution to  \eqref{Mestimator} with 
\textcolor{blue}{\begin{equation}\label{MLT}
u(d;\nu)=\frac{2p+\nu}{\nu+2d}.
\end{equation}}
 The choice of $\nu$ closely relates to the heavy-tailedness of the data set, T-distribution with small $\nu$ corresponding to heavier tailed distribution than it with large $\nu$. For $\nu \to \infty$, t-distribution degenerates to Gaussian distribution, and the estimator becomes SCM, which has no ability to suppress extreme data points. However, in the spectrum sensing problem, we wish to achieve robustness, so we prefer to choose a small value such as $\nu \leq 5$. This estimator can also degenerate to Tyler's M-estimator by setting $\nu=0$, which does not correspond to any natural t-distribution. This estimator is an intermediate estimator between SCM and Tyler's M-estimator. This estimator is computed by \eqref{iter1}.
\section{Robust Eigenvalue Based Spectrum Sensing}
The spectrum sensing with single source setting is considered, where each SUs equipped with $p$ antennas and the test statistics is computed based $n$ time samples.

The simplest version of the spectrum sensing is the detection of a signal from a noisy environment. This task can be formulated by a hypothesis test, whose null hypothesis is that a signal does not exist, and the alternative hypothesis is that a signal exists. The received signal samples under two hypothesis are formulated as,
    \begin{equation} \label{Hypothesis}
   \boldsymbol{x} (i)=\left\{
        \begin{array}{ll}
            \boldsymbol{z}(i) & \quad  H_0:\textnormal{signal does not exist} \\
            s(i)\boldsymbol{h}+\boldsymbol{z}(i) & \quad H_1:\textnormal{signal exists},
        \end{array}
    \right.
    \end{equation} 
where $\boldsymbol{x}(i)\in \mathbb{C}^p$ is the received sample vector at instant $i$ of one SU, $\boldsymbol{h} \in \mathbb{C}^p$ represents the fading channel, $s(i) \in \mathbb{C}$ is the transmitted symbol modeled as a complex Gaussian random variable with zero mean and unit variance, and $\boldsymbol{z}(i) \in \mathbb{C}^p$ is the received noise vector which is assumed to be i.i.d in time, with mean zero and covariance $\boldsymbol{\sigma}^2\boldsymbol{I}$ not necessarily Gaussian distributed. We assume the channel $\boldsymbol{h}$ being constant during $i=1,\dots ,n$ transmissions. Under $H_0$, the received sample is pure noise whose population covariance matrix is $E[\boldsymbol{x}(i)\boldsymbol{x}(i)^H] =\boldsymbol{\sigma}^2\boldsymbol{I}$ and the largest eigenvalue of the population covariance is $\sigma^2$. Under $H_1$, the received sample is the noise plus signal, whose population covariance matrix is $E[\boldsymbol{x}(i)\boldsymbol{x}(i)^H]=\boldsymbol{h}\boldsymbol{h}^H+\sigma^2\boldsymbol{I}$ and the largest eigenvalue of the population covariance is $\|\boldsymbol{h}\|^2+\sigma^2$. Also we define the signal to noise ratio(SNR) at the receiver as,
\begin{equation}
\rho = \frac{E\|\boldsymbol{h}s(i)\|^2}{E\|\boldsymbol{z}(i)\|^2}=\frac{E\|\boldsymbol{h}\|^2}{p\sigma^2}.
\end{equation}

The received sample matrix generated by the system is a $p \times n$ matrix consisting of all the sample vectors from $p$ antennas:
\[
\boldsymbol{X} =
\begin{bmatrix}
    x_1{(1)} & x_1{(2)} & x_1{(3)} & \dots  & x_1{(n)} \\
    x_2{(1)} & x_2{(2)} & x_2{(3)} & \dots  & x_2{(n)} \\
    \vdots & \vdots & \vdots & \ddots & \vdots \\
    x_p{(1)} & x_p{(2)} & x_p{(3)} & \dots  & x_p{(n)}
\end{bmatrix}.
\]
The SCM $\boldsymbol{S}$ is 
\begin{equation} \label{SCM}
\boldsymbol{S}=\frac{1}{n}\boldsymbol{X}\boldsymbol{X}^H.
\end{equation}
The Tyler's M-estimator $\boldsymbol{\hat{\Sigma}}_{TY}$ is
\begin{equation} \label{TY}
\boldsymbol{\hat{\Sigma}}_{TY}=\frac{p}{n}\sum_{i=1}^n \frac{\boldsymbol{x}_i\boldsymbol{x}_i^H}{\boldsymbol{x}_i^H\boldsymbol{\hat{\Sigma}}_{TY}^{-1}\boldsymbol{x}_i}.
\end{equation} 
Let $\lambda^{S}_1\geq\dots\geq\lambda^{S}_p$ and $\lambda^{TY}_1\geq\dots\geq\lambda^{TY}_p$ be the eigenvalues of $S$ and  $\hat{\Sigma}_{TY}$ respectively.

In general, let $T$ be the test statistic employed by the detector to distinguish between $H_0$ and $H_1$. The detector makes the decision by comparing the test statistics $T$ computed from the data with a pre-determined threshold $t$: if $T>t$ it decides that $H_1$ is true, otherwise $H_0$ is true. The performance of spectrum sensing can be primarily determined based on two metrics: the probability of detection (POD) and the probability of false alarm (POF). POD is defined as
$$P_d=Pr(T>t|H_1),$$ and POF is defined as $$P_{fa}=Pr(T>t|H_0).$$ POD is closely related to quality-of-service (QoS) of PUs since low POD means that the communication of PUs will be interfered often by SUs. POF is closely associated with the QoS of SUs since a false alarm will reduce the spectral usage efficiency. The optimal detector for spectrum sensing usually has the maximized POD given the constraint of the POF.

When the noise vector is Gaussian distributed, there are two nearly optimal test statistics, i.e. Roy's largest root test (RLRT) and a generalized likelihood ratio test (GLRT). The RLRT requires the knowledge of noise power while GLRT does not require such knowledge. The RLRT asymptotically determines the Neyman-Pearson(NP) likelihood ratio \cite{muirhead1978latent,kritchman2009non}which gives the most powerful test in the case of a simple hypothesis test. The RLRT statistics is defined as,
\begin{equation} \label{RLRT}
T^S_{RLRT}=\frac{\lambda^S_1}{\sigma^2}.
\end{equation} 
When the noise power is unknown, the hypothesis test becomes a composite hypothesis test ,and the NP likelihood ratio is not available. A common procedure is the generalized likelihood ratio test which in our model is \cite{bianchi2011performance}
\begin{equation} \label{GLRT}
T^S_{GLRT}=\frac{\lambda^S_1}{\frac{1}{p}{\tr}(\boldsymbol{S})}.
\end{equation}

Those test statistics derived from the SCM preserves certain optimality when the noise vector is Gaussian. When the noise vector is CES distributed with heavy tails, those test statistics will lose their optimality and have very high variance, i.e. with high probability the statistics are far away from their population counterparts. In the hypothesis test, the SCM based detector tends to confuse signal transmitted by PUs and the effect of impulsive effect which leads to a high POF given a fixed POD. To deal with the deficiency of SCM, we derive these two statistics from $\boldsymbol{\hat{\Sigma}}_{TY}$. The proposed test statistics will be 
\begin{equation}
T^{TY}_{RLRT}=\frac{\lambda^{TY}_1}{\sigma^2}.
\end{equation} 
and 
\begin{equation}
T^{TY}_{GLRT}=\frac{\lambda^{TY}_1}{\frac{1}{p}{\tr}(\boldsymbol{\hat{\Sigma}_{TY}})}.
\end{equation} 
Similar to $T^S_{SCM}$, the detector using the latter statistics requires no knowledge of noise power. These two statistics can also be derived from other M-estimators by choosing different $u$ functions like \eqref{MLT}. However many of those choices have free parameters to adjust according to the noise environment ,which requires certain amount of data samples to learn the noise first but in cognitive radio applications the time slot to sensing the spectrum is limited.

There are several reasons to use Tyler's M-estimator other than other M-estimators when the noise is CES distributed. Firstly, this estimator cancels out the effect of texture parameter shown in (\ref{CES}) which means the behavior of this estimator and functions of this estimator do not relate to the exact noise distribution if the data is CES distributed. Then the statistics derived from the Tyler's M-estimator have a constant POF under a broad class of data distribution with respect to a given threshold $t$. \textcolor{blue}{In addition, $T^{TY}_{GLRT}$ and $T^{TY}_{RLRT}$ have the same performance. This can be explained by the fact that the ratio of these two statistics, $\frac{\sigma}{{\tr}(\boldsymbol{\hat{\Sigma}_{TY}})}=\frac{\sigma}{\alpha}$, is a constant under any hypothesis (\ref{iter}). However, the ratio derived from the SCM is not the same in different realizations, thus they have different performance.} Secondly, the performances of those tests are better than those derived from the SCM under heavy-tailed data. Last but not least, this estimator does not need to learn the data distribution non-parametrically in order to optimize its performance when the sample size $n$ is limited. If the available sample size $n$ is large, $\hat{\Sigma}_M$ can be used with the parameter $\nu$ optimized. If the available sample size is large enough to learn the noise distribution, the ML-estimator may be the best choice.

\section{Simulation and Numerical Result}
\begin{figure}
\centering
\includegraphics[width=.75\linewidth]{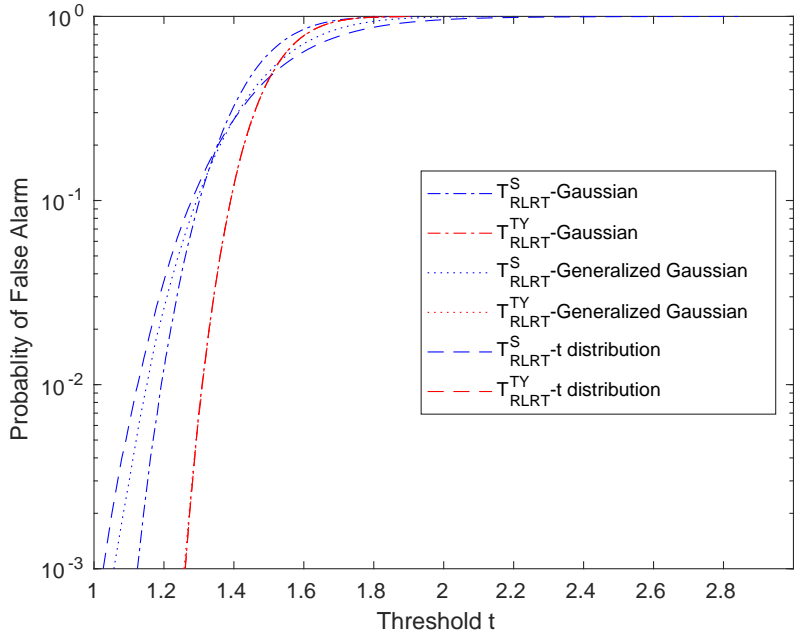}
\caption{Constant POF property for $T^{TY}_{RLRT}$ when the noise power is known}
\label{CPOF1}
\end{figure}

\begin{figure}
\centering
\includegraphics[width=.75\linewidth]{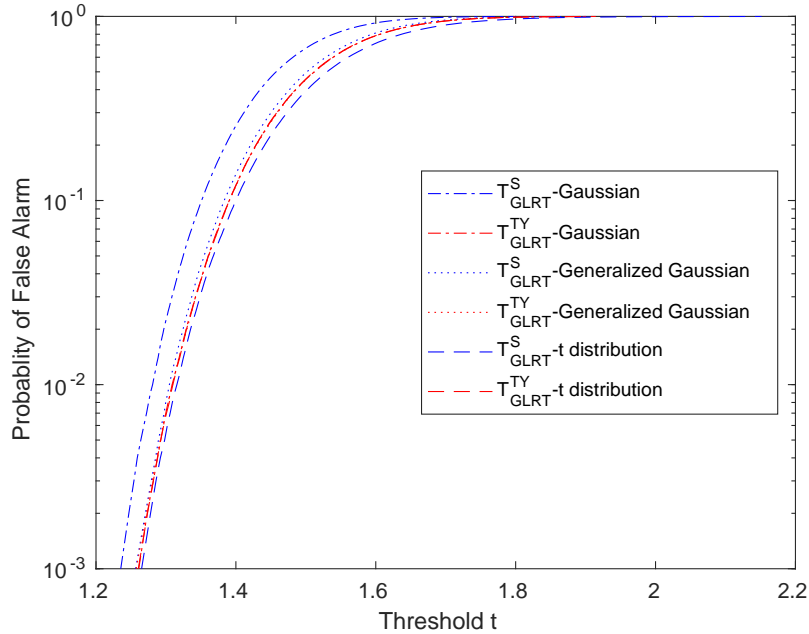}
\caption{Constant POF property for $T^{TY}_{GLRT}$ when the noise power is unknown}
\label{CPOF2}
\end{figure}

Figure \ref{CPOF1} and \ref{CPOF2} illustrate the constant POF property for $T^{TY}_{RLRT}$ and $T^{TY}_{GLRT}$ with regard to different CES distributions. The figures represent the relation between the threshold $t$ and the POF $P_{fa}$ for different CES distributions: Gaussian, Generalized Gaussian, and Student-t. We generated those four test statistics $100000$ times for noise data from these three noise distributions respectively with $n=10$ and $p=5$. With those generated data we computed empirical cumulative distributions of those statistics which are Figure \ref{CPOF1} and \ref{CPOF2}. We can notice that the statistics derived from the SCM have different curves when the noise distribution changes while the statistics derived from the $\boldsymbol{\hat{\Sigma}}_{TY}$ always have the same curve. Thus, we conclude that distributions of $T^{TY}_{RLRT}$ and $T^{TY}_{GLRT}$ do not change in different CES noise. This property enables us to derive a constant probability of false alarm test based on $T^{TY}_{RLRT}$ and $T^{TY}_{GLRT}$ in all CES distributions.


\begin{figure} 
\centering
\includegraphics[width=.75\linewidth]{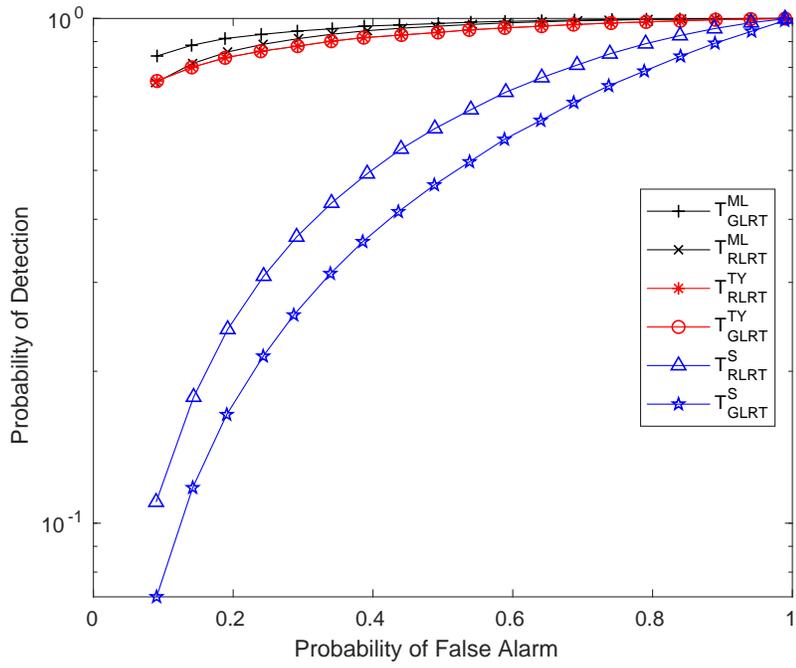}
\caption{Performance of the proposed detector under Generalized Gaussian Noise}
\label{ROC}
\end{figure}

In figure \ref{ROC} we have receiver's operation curves for different test statistics under impulsive noise. Each simulation was repeated $100000$ times for $n=50$, $p=5$ and $\rho=0dB$. The simulation results compare the performance of different tests under Generalized Gaussian noise with $s=0.1$ \cite{ollila2012complex}. As reference, we also have $T^{ML}_{RLRT}$ and $T^{ML}_{GLRT}$ derived from the ML-estimator for Generalized Gaussian Distribution. The ML-estimator is \eqref{MLestimator} with $u(d)=\frac{s}{b}d^{s-1}$, where $b=[p\Gamma(\frac{p}{s})/\Gamma(\frac{p+1}{s})]^s$. Those test statistics have the best performance but require exact knowledge of the noise distribution ,which is usually unavailable in practice. The performance of $T^{TY}_{RLRT}$ and $T^{TY}_{GLRT}$ are exactly the same and outperform both $T^{S}_{RLRT}$ and $T^{S}_{GLRT}$ in the impulsive noise environment. The gap between detectors using $\boldsymbol{\hat{\Sigma}}_{TY}$ and the detectors using ML-estimator is not significant. The gap can be interpreted as the price paid for the robustness we gained from using $\boldsymbol{\hat{\Sigma}}_{TY}$. $T^{S}_{RLRT}$ with knowledge of the noise power outperforms $T^{S}_{GLRT}$ as expected. If the sample size $n$ is large, tests based on $\boldsymbol{\hat{\Sigma}}_M$ can perform no worse than tests based on $\boldsymbol{\hat{\Sigma}}_{TY}$ since the parameter $\nu$ can be set to be $0$ or another optimized value with proper a optimization algorithm. 

\begin{figure} 
\centering
\includegraphics[width=.75\linewidth]{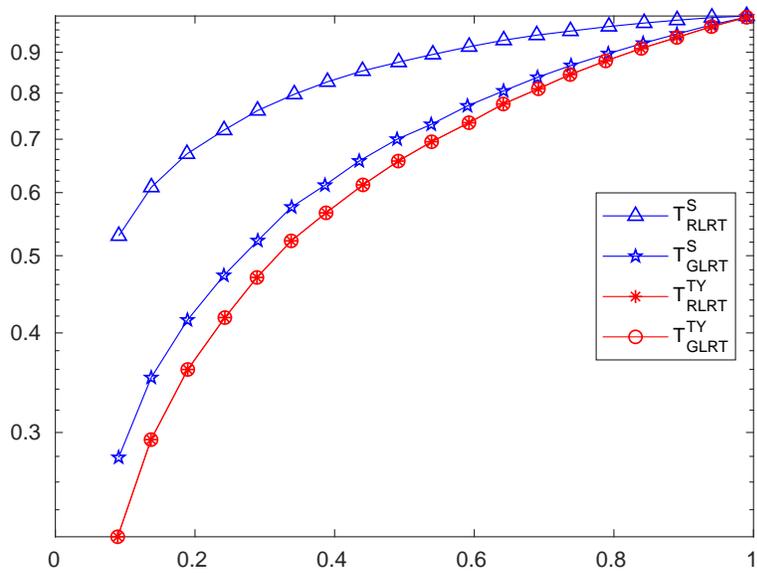}
\caption{Performance of the proposed detector under Gaussian Noise}
\label{ROCGaussian}
\end{figure}

In figure \ref{ROCGaussian} we have receiver's operation curves for test statistics derived from $S$ and $\boldsymbol{\hat{\Sigma}}_{TY}$ under Gaussian noise. This simulation is implemented with the same setting as in the previous figure except for the noise type. The simulation results show the performance loss by using robust estimator in stead of SCM in Gaussian noise. The detector using $T^{S}_{RLRT}$ has the best performance, but in practice this statistic is usually unavailable since the noise power is not known. The gap between robust detectors and $T^{S}_{GLRT}$ is not big in the simulation. This demonstrates the price pay for robustness is not high in Gaussian noise as well.

\section{Conclusion}
A blind robust eigenvalue-based detection has been proposed in this paper ,which is insensitive to CES distribution and noise power. The constant probability of false alarm regards to different type of CES distribution has been shown numerically. In addition, the robustness of this detector is shown numerically in both Gaussian and non-Gaussian noise environment.

Based on these results, further study should be done to derive a closed-form expression of probability of false alarm and probability of detection used to design the detector accurately. Also, a proper optimization procedure can be proposed to optimize $\boldsymbol{\hat{\Sigma}}_M$ in the case of sufficiently large sample size $n$.

\bibliography{citation}

\begin{thebibliography}{10}

\bibitem{cognitivesurvey}
Ian~F. Akyildiz, Won-Yeol Lee, Mehmet~C. Vuran, and Shantidev Mohanty.
\newblock Next generation/dynamic spectrum access/cognitive radio wireless
  networks: A survey.
\newblock {\em Computer Networks}, 50(13):2127 -- 2159, 2006.

\bibitem{wang2014spatial}
Wang Jinlong, Ding Guoru, Wu~Qihui, Shen Liang, and Song Fei.
\newblock Spatial-temporal spectrum hole discovery: a hybrid spectrum sensing
  and geolocation database framework.
\newblock {\em Chinese Science Bulletin}, 59(16):1896--1902, 2014.

\bibitem{cosensingRMTGLRT}
P.~Bianchi, J.~Najim, G.~Alfano, and M.~Debbah.
\newblock Asymptotics of eigenbased collaborative sensing.
\newblock In {\em 2009 IEEE Information Theory Workshop}, pages 515--519, Oct
  2009.

\bibitem{cosensingRMTeigen}
Y.~Zeng and Y.~. Liang.
\newblock Eigenvalue-based spectrum sensing algorithms for cognitive radio.
\newblock {\em IEEE Transactions on Communications}, 57(6):1784--1793, June
  2009.

\bibitem{cosensingRMT}
L.~S. Cardoso, M.~Debbah, P.~Bianchi, and J.~Najim.
\newblock Cooperative spectrum sensing using random matrix theory.
\newblock In {\em 2008 3rd International Symposium on Wireless Pervasive
  Computing}, pages 334--338, May 2008.

\bibitem{urkowitz1967energy}
Urkowitz Harry.
\newblock Energy detection of unknown deterministic signals.
\newblock {\em Proceedings of the IEEE}, 55(4):523--531, 1967.

\bibitem{poor2013introduction}
Poor~H Vincent.
\newblock {\em An introduction to signal detection and estimation}.
\newblock Springer Science \& Business Media, 2013.

\bibitem{enserink1994cyclostationary}
Enserink Scott and Cochran Douglas.
\newblock A cyclostationary feature detector.
\newblock In {\em Proceedings of 1994 28th Asilomar Conference on Signals,
  Systems and Computers}, volume~2, pages 806--810. IEEE, 1994.

\bibitem{middleton1999non}
Middleton David.
\newblock Non-gaussian noise models in signal processing for
  telecommunications: new methods an results for class a and class b noise
  models.
\newblock {\em IEEE Transactions on Information Theory}, 45(4):1129--1149,
  1999.

\bibitem{batur2008measurements}
Batur~Okan Z, Koca Mutlu, and Dundar Gunhan.
\newblock Measurements of impulsive noise in broad-band wireless communication
  channels.
\newblock In {\em 2008 Ph. D. Research in Microelectronics and Electronics},
  pages 233--236. IEEE, 2008.

\bibitem{taher2008microwave}
Taher~Tanim M, Misurac~Matthew J, LoCicero~Joseph L, and Ucci~Donald R.
\newblock Microwave oven signal interference mitigation for wi-fi communication
  systems.
\newblock In {\em 2008 5th IEEE Consumer Communications and Networking
  Conference}, pages 67--68. IEEE, 2008.

\bibitem{blankenship1997measurements}
Blankenship~T Keith, Kriztman DM, and Rappaport~Theodore S.
\newblock Measurements and simulation of radio frequency impulsive noise in
  hospitals and clinics.
\newblock In {\em 1997 IEEE 47th Vehicular Technology Conference. Technology in
  Motion}, volume~3, pages 1942--1946. IEEE, 1997.

\bibitem{lunden2010robust}
Lund{\'e}n Jarmo, Kassam~Saleem A, and Koivunen Visa.
\newblock Robust nonparametric cyclic correlation-based spectrum sensing for
  cognitive radio.
\newblock {\em IEEE Transactions on Signal Processing}, 58(1):38--52, 2010.

\bibitem{kang2010class}
Kang~Hyun Gu, Song Iickho, Yoon Seokho, and Kim~Yun Hee.
\newblock A class of spectrum-sensing schemes for cognitive radio under
  impulsive noise circumstances: Structure and performance in nonfading and
  fading environments.
\newblock {\em IEEE Transactions on Vehicular Technology}, 59(9):4322--4339,
  2010.

\bibitem{moghimi2011adaptive}
Moghimi Farzad, Nasri Amir, and Schober Robert.
\newblock Adaptive l\_p—norm spectrum sensing for cognitive radio networks.
\newblock {\em IEEE Transactions on Communications}, 59(7):1934--1945, 2011.

\bibitem{wimala2011}
Wimalajeewa Thakshila and Varshney~Pramod K.
\newblock Polarity-coincidence-array based spectrum sensing for multiple
  antenna cognitive radios in the presence of non-gaussian noise.
\newblock {\em IEEE Transactions on Wireless Communications}, 10(7):2362--2371,
  2011.

\bibitem{ollila2012complex}
Esa Ollila, David~E Tyler, Visa Koivunen, and H~Vincent Poor.
\newblock Complex elliptically symmetric distributions: Survey, new results and
  applications.
\newblock {\em IEEE Transactions on signal processing}, 60(11):5597--5625,
  2012.

\bibitem{maronna2018robust}
Ricardo~A Maronna, R~Douglas Martin, Victor~J Yohai, and Mat{\'\i}as
  Salibi{\'a}n-Barrera.
\newblock {\em Robust Statistics: Theory and Methods (with R)}.
\newblock Wiley, 2018.

\bibitem{maronna1976robust}
Ricardo~Antonio Maronna.
\newblock Robust m-estimators of multivariate location and scatter.
\newblock {\em The annals of statistics}, pages 51--67, 1976.

\bibitem{kent1991redescending}
John~T Kent, David~E Tyler, et~al.
\newblock Redescending $ m $-estimates of multivariate location and scatter.
\newblock {\em The Annals of Statistics}, 19(4):2102--2119, 1991.

\bibitem{ollila2003robust}
Esa Ollila and Visa Koivunen.
\newblock Robust antenna array processing using m-estimators of
  pseudo-covariance.
\newblock In {\em 14th IEEE Proceedings on Personal, Indoor and Mobile Radio
  Communications, 2003. PIMRC 2003.}, volume~3, pages 2659--2663. IEEE, 2003.

\bibitem{kent1997data}
John~T Kent.
\newblock Data analysis for shapes and images.
\newblock {\em Journal of statistical planning and inference}, 57(2):181--193,
  1997.

\bibitem{muirhead1978latent}
Robb~J Muirhead.
\newblock Latent roots and matrix variates: a review of some asymptotic
  results.
\newblock {\em The Annals of Statistics}, pages 5--33, 1978.

\bibitem{kritchman2009non}
Shira Kritchman and Boaz Nadler.
\newblock Non-parametric detection of the number of signals: Hypothesis testing
  and random matrix theory.
\newblock {\em IEEE Transactions on Signal Processing}, 57(10):3930--3941,
  2009.

\bibitem{bianchi2011performance}
Pascal Bianchi, Merouane Debbah, Myl{\`e}ne Ma{\"\i}da, and Jamal Najim.
\newblock Performance of statistical tests for single-source detection using
  random matrix theory.
\newblock {\em IEEE Transactions on Information theory}, 57(4):2400--2419,
  2011.

\end{thebibliography}
\bibliographystyle{unsrt}

\end{document}